\documentclass[preprint,onecolumn]{aastex}

\usepackage{graphicx}
\usepackage{amsmath}
\usepackage{amssymb}
\usepackage{units}
\usepackage{color}
\usepackage{hyperref}
\usepackage{lineno}
\usepackage{natbib}

\begin{document}

\title{Lowering IceCube's Energy Threshold for Point Source Searches in the Southern Sky}

\author{
IceCube Collaboration:
M.~G.~Aartsen\altaffilmark{1},
K.~Abraham\altaffilmark{2},
M.~Ackermann\altaffilmark{3},
J.~Adams\altaffilmark{4},
J.~A.~Aguilar\altaffilmark{5},
M.~Ahlers\altaffilmark{6},
M.~Ahrens\altaffilmark{7},
D.~Altmann\altaffilmark{8},
K.~Andeen\altaffilmark{9},
T.~Anderson\altaffilmark{10}
% ...
}

% ...
\altaffiltext{1}{Department of Physics, University of Adelaide, Adelaide, 5005, Australia}
\altaffiltext{2}{Physik-department, Technische Universit\"at M\"unchen, D-85748 Garching, Germany}
\altaffiltext{3}{DESY, D-15735 Zeuthen, Germany}
\altaffiltext{4}{Dept.~of Physics and Astronomy, University of Canterbury, Private Bag 4800, Christchurch, New Zealand}
\altaffiltext{5}{Universit\'e Libre de Bruxelles, Science Faculty CP230, B-1050 Brussels, Belgium}
\altaffiltext{6}{Dept.~of Physics and Wisconsin IceCube Particle Astrophysics Center, University of Wisconsin, Madison, WI 53706, USA}
\altaffiltext{7}{Oskar Klein Centre and Dept.~of Physics, Stockholm University, SE-10691 Stockholm, Sweden}
\altaffiltext{8}{Erlangen Centre for Astroparticle Physics, Friedrich-Alexander-Universit\"at Erlangen-N\"urnberg, D-91058 Erlangen, Germany}
\altaffiltext{9}{Department of Physics, Marquette University, Milwaukee, WI, 53201, USA}
\altaffiltext{10}{Dept.~of Physics, Pennsylvania State University, University Park, PA 16802, USA}
% ...

\begin{abstract}
Observation of a point source of astrophysical neutrinos would be a ``smoking gun" signature of a cosmic-ray accelerator. While IceCube has recently discovered a diffuse flux of astrophysical neutrinos, no localized point source has been observed. Previous IceCube searches for point sources in the southern sky were restricted by either an energy threshold above a few hundred \unit{TeV} or poor neutrino angular resolution. Here we present a search for southern sky point sources with greatly improved sensitivities to neutrinos with energies below \unit[100]{TeV}.  By selecting charged-current $\nu_{\mu}$ interacting inside the detector, we reduce the atmospheric background while retaining efficiency for astrophysical neutrino-induced events reconstructed with sub-degree angular resolution. The new event sample covers three years of detector data and leads to a factor of ten improvement in sensitivity to point sources emitting below \unit[100]{TeV} in the southern sky. No statistically significant evidence of point sources was found, and upper limits are set on neutrino emission from individual sources. A posteriori analysis of the highest-energy ($\sim$\unit[100]{TeV}) starting event in the sample found that this event alone represents a $2.8\sigma$ deviation from the hypothesis that the data consists only of atmospheric background.
\end{abstract}
\maketitle



\section{Introduction}

Despite 100 years of cosmic-ray observations, the sources of the highest energy cosmic rays remain unknown~\citep{BeattyWesterhoff,KoteraOlinto}. Cosmic rays are predicted to be accelerated to very high energies in astrophysical objects, where they produce neutrinos upon interacting with matter or photons~\citep{Stecker,WaxmanBahcall,LearnedMannheim,AnchoReview}. The properties of neutrinos make them a unique astrophysical messenger. Interacting only via the weak force, they can travel astronomical distances without experiencing significant absorption. Neutral in charge, neutrinos follow straight paths through space even in magnetic fields, thus pointing back to their origin. Since high-energy cosmic neutrinos are only known to be produced in interactions of accelerated hadrons, observing neutrino point sources would identify cosmic-ray sources.


The IceCube Neutrino Observatory observed an excess of high-energy neutrinos that is consistent with a diffuse astrophysical neutrino flux \citep{HESESciencePaper,HESEPRL,JakobCascades,WeaverNuMu}. The origin of this flux remains unknown, and no significant clustering or correlation among the highest-energy events has been observed. Additional searches for point-source emission by IceCube and ANTARES using muon tracks have found neither evidence for point-like nor extended sources~\citep{4YrPSPaper,IC86TimeDependentPaper,ANTARESPS}. A number of explanations for this flux have been proposed; see \citet{AnchoReview} and references therein. 

Potential sources of high-energy neutrinos include supernova remnants (SNR) \citep{Cavasinni2006,HalzenGC,KistlerBeacom}, pulsars~\citep{BednarekProtheroe97,LinkBurgio1,KeFang12}, active galactic nuclei (AGN) \citep{Kalashev2013PRL,Murase2014AGN,Stecker} and starburst galaxies \citep{2011ApJStarbursts,LoebWaxman,2003ApJStarbursts}. Many potential Galactic sources are in the southern sky and are predicted to accelerate cosmic rays to \unit{PeV} energies, producing $\unit{TeV} - \unit{PeV}$ neutrinos. Additionally, gamma-ray telescopes observe many Galactic sources with energy cutoffs below \unit[100]{TeV}~\citep{FermiSNR,HESSRXJ17}, suggesting Galactic neutrino emission may be most prominent in this energy range.


Southern sky point source searches with IceCube are difficult at these energies because atmospheric muons are backgrounds in this region of sky, triggering the detector at a rate of \unit[2.5]{kHz}. They have a softer energy spectrum than the expected signal, and can be reduced by selecting well-reconstructed high-energy throughgoing tracks~\citep{PSDowngoing,4YrPSPaper}. This strategy increases the effective detector volume with neutrino-induced muons originating outside the detector, which can be reconstructed to \textless $1^{\circ}$. However, the large background requires a high energy threshold, reducing the sensitivity below \unit[1]{PeV}. An alternative strategy removes the atmospheric muon background by selecting high-energy contained-vertex events---bright events that start inside the detector~\citep{HESESciencePaper,HESEPRL,JakobCascades}. This removes the majority of the background, providing a signal-dominated sample. However, most of these events are spherical light deposition from charged-current $\nu_e$ or $\nu_\tau$ interactions or all-flavor neutral-current interactions. These events have $\sim15^{\circ}$ angular resolutions, restricting their utility for point source searches.

The analysis presented here lowers IceCube's energy threshold in the southern sky by selecting charged-current $\nu_{\mu}$ events that start inside the detector. Compared to the high-energy contained-vertex event search~\citep{HESEPRL}, this enhances the $\nu_{\mu}$ effective area below $\sim\unit[200]{TeV}$, thereby increasing the expected rate of signal events with $<1^{\circ}$ angular resolution. Compared to the throughgoing muon analysis~\citep{4YrPSPaper}, the background rate is reduced by two orders of magnitude while retaining similar angular resolution, energy resolution, and neutrino effective area below $\sim\unit[100]{TeV}$. This selection provides a nearly independent event sample, which is combined with the throughgoing muon data in a joint likelihood fit.



\section{Event Selection and Analysis Technique}

The IceCube Neutrino Observatory is a cubic-kilometer array of photomultiplier tubes (PMTs) embedded in the glacial ice at the geographic South Pole~\citep{DAQPaper,FirstYearPerformancePaper}. Eighty-six cables (called strings) are instrumented with 5160 PMTs, \unit[$1.5 - 2.5$]{km} below the surface of the glacier. Neutrinos interact in the ice and produce charged leptons that are detected via their Cherenkov radiation by the PMTs~\citep{PMTPaper}. The analysis presented here searches for ``starting tracks"---charged-current $\nu_\mu$ events characterized by a large initial energy loss inside the detector followed by a track pattern as the resulting muon traverses and exits the detector. The dominant background for this search consists of atmospheric muons produced in air showers, a small fraction of which deposit little light upon entering the detector and therefore pass the starting track event selection. Atmospheric neutrinos form a subdominant background but are indistinguishable from astrophysical neutrinos, unless a muon from the same air shower also reaches the detector~\citep{JakobVetoPaper,SchonertVetoPaper}.


The starting track event selection is applied to data collected between 2010 and 2013. This includes one year of data from the 79-string detector configuration and two years from the completed 86-string detector. The event selection is performed using a two-stage veto. First, events with hits on the outer layer of the detector are removed using the veto algorithm from~\cite{HESESciencePaper}. The veto region includes the outer strings, the top \unit[90]{m} and bottom \unit[10]{m} of the detector, and a \unit[60]{m} layer of PMTs in a region of dusty ice in the middle of the detector. Any event with more than two of its first 250 observed photoelectrons (PE) in the veto region is removed. Additionally, events with fewer than 1500 total PEs are removed. This leaves $\sim 3900$ events per year, mostly atmospheric muons. In contrast, the search in~\cite{HESESciencePaper} used the same veto criteria, but required events to deposit 6000 PE or greater, and found 37 events in the same dataset. 

Most remaining events still start closer to the border of the detector than expected for a collection of true starting event. This is an energy-dependent effect---high-energy background is more likely to emit observable light near the border, while low-energy background events can pass by a couple of strings without being detected. Two likelihood-based reconstructions are used. One estimates the position of the first energy loss by fitting for positions of stochastic losses along the track, and the other fits the muon energy loss along its track as a proxy for the total muon energy~\citep{EnergyRecoPaper}. A two-dimensional cut on these quantities removes 95\% of the remaining background while retaining 91\% of simulated $\nu_{\mu}$ drawn from an $E^{-2}$ spectrum.

Additionally, poorly-reconstructed events and upgoing events are removed. Events with unreliable directions, where two algorithms have poor agreement, cannot help identify point sources and are removed. Starting events in the upgoing region (zenith angle $>85^{\circ}$) are not included because the Earth and ice overburden in this range allow a sufficiently pure neutrino sample to be obtained with through-going events, as in the conventional point source samples~\citep{4YrPSPaper}. After all cuts, the three-year event sample contains 549 events.



We search for point sources using the un-binned maximum likelihood method from~\citet{IC79PSPaper,4YrPSPaper}:

\begin{equation}
\label{eq:likelihoodMESE}
\mathcal{L}(\gamma, n_{s}) = \prod_{j} \mathcal{L}^{j} (\gamma, n^{j}_{s}) = \prod_{j}\;\prod_{i \in j} \left[\frac{n^{j}_{s}}{N^{j}}\mathcal{S}^{j}_{i} + \left(1- \frac{n^{j}_{s}}{N^{j}}\right)\mathcal{B}^{j}_{i} \right].
\end{equation}
The first product is over the different datasets $j$, and $i \in j$ indicates the $i$th event belongs to the $j$th dataset. We combine the three-year starting track sample with the four-year throughgoing muon sample~\citep{4YrPSPaper}, which includes data from partial IceCube configurations during construction. The number of neutrino events originating from the point source, $n_s$, and the spectral index for a source with a power-law spectrum, $\gamma$, are free parameters in the fit. The number of source events is the sum of the source events fitted within each dataset, $n^{j}_{s}$. The ratios of the $n^{j}_{s}$ are fixed by the relative number of signal events expected in each dataset for the hypothesis being tested (point source at declination $\delta$ with spectral index $\gamma$). The probability distribution functions (PDFs) $\mathcal{S}^{j}_{i}$ and $\mathcal{B}^{j}_{i}$ describe the spatial and energy distributions of the signal and background, respectively. Since our event sample mostly contains tracks, we apply the same energy and directional reconstructions used in~\cite{4YrPSPaper}, resulting in a median angular resolution below one degree for neutrinos with energies above \unit[65]{TeV}.


We perform two hypothesis tests to search for point sources. The first is an all-sky likelihood scan, where the likelihood (Eq.~\ref{eq:likelihoodMESE}) is maximized independently at each location in the sky on a $0.1^{\circ} \times 0.1^{\circ}$ grid. The final results of this test are the location, best-fit parameters, and $p$-value of the most significant excess (the ``hottest spot"). This is identical to~\cite{4YrPSPaper}, except we restrict our search for the hottest spot to declinations between -85$^{\circ}$ and -5$^{\circ}$. The chance probability of finding a hottest spot as significant as the one observed is estimated by repeating the test on an ensemble of datasets randomized in right ascension.

To reduce the large number of effective trials associated with scanning the entire sky, the second hypothesis test searches for neutrino emission from a catalog of candidate sources. These sources are selected based on multi-wavelength observations or astrophysical models predicting neutrino emission. We apply the same catalog used in the southern sky in~\cite{HESEPRL}, which contains 38 {\it a priori}-selected sources from previous IceCube and ANTARES analyses~\citep{4YrPSPaper,ANTARESPS}. Similar to the all-sky search, the final post-trial $p$-value is obtained by repeating this test on randomized datasets.

\begin{figure}
\centering
\includegraphics[width=0.7\linewidth]{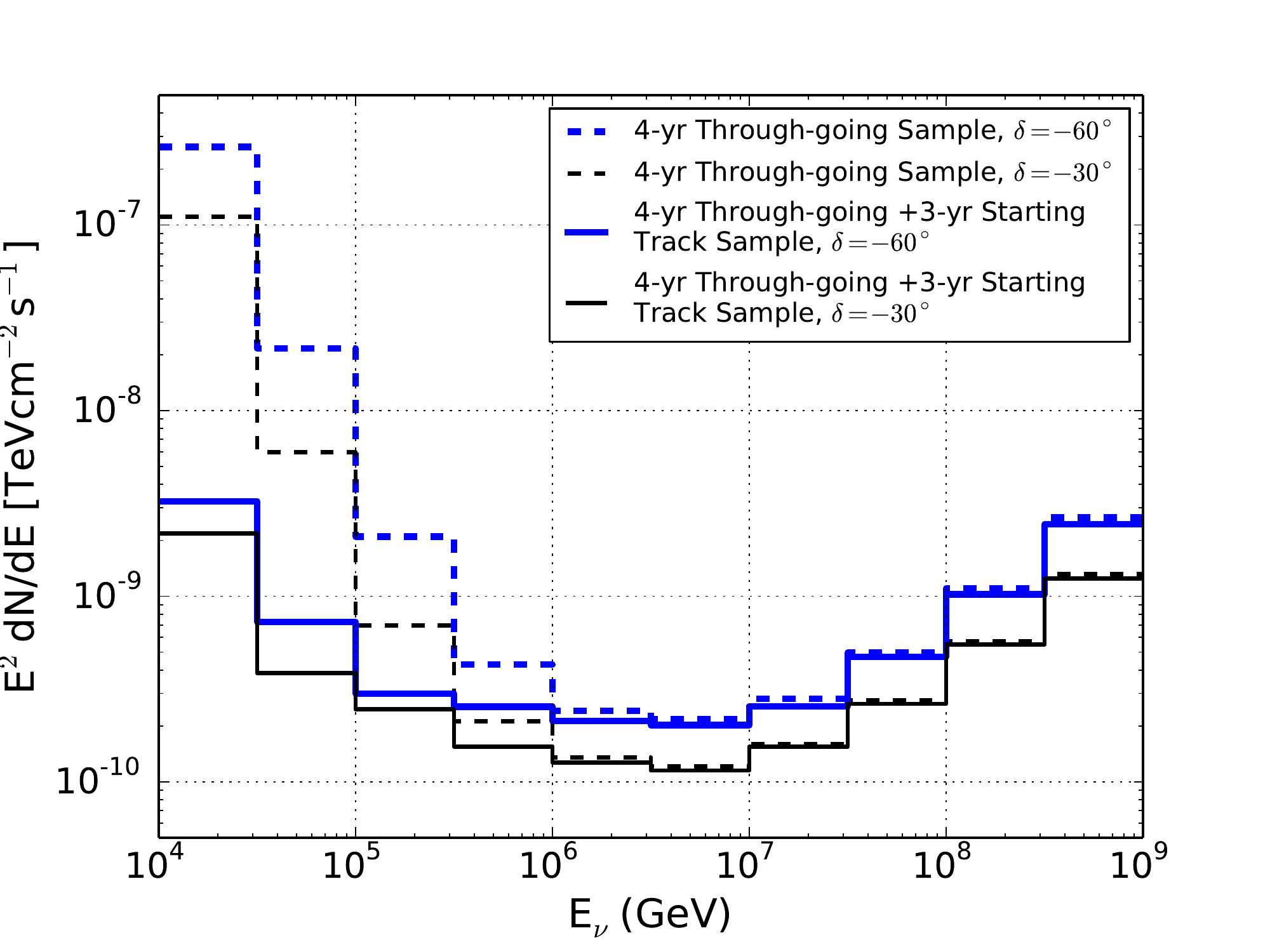}
\caption[Discovery flux as a function of neutrino energy]{Discovery flux as a function of neutrino energy at $5\sigma$ confidence level. Point sources with an E$^{-2}$ spectrum are simulated over a half-decade in energy, and the flux in each bin required for discovery forms the curve above. Previous IceCube results~\citep{4YrPSPaper} are shown with dashed lines. The new event selection lowers the discovery threshold at \unit[$10^5$]{GeV} by a factor of $\sim10$.}
\label{fig:DiffDisco}
\end{figure}

Including the starting track sample leads to a substantial improvement in the discovery potential for point sources (Figure~\ref{fig:DiffDisco}). For southern sky sources with a cutoff at \unit[1]{PeV}, the discovery potential is a factor of $2-3$ times better than previous results~\citep{4YrPSPaper}. For sources with a cutoff at \unit[100]{TeV}, the improvement is a factor of 10.

\section{Results}

\begin{figure}[t!]
\hspace{-2cm}  
\includegraphics[width=1.25\textwidth]{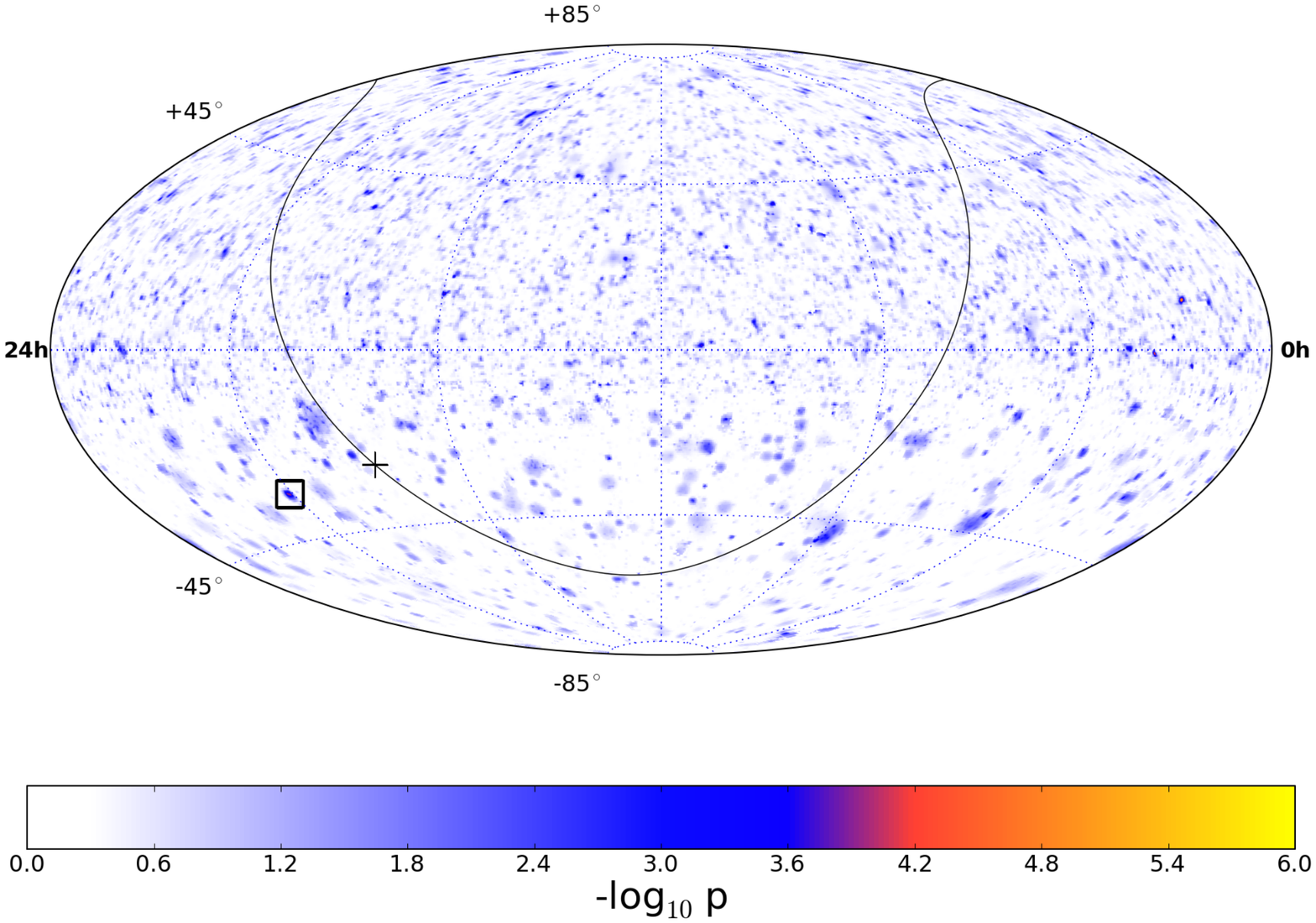}
\vspace{-1cm}
\caption[Pre-trial significance skymap of the starting track point source scan]{Pre-trial significance skymap in equatorial coordinates (J2000) for the starting track sample combined with the throughgoing muon sample. The black line indicates the Galactic Plane, and the black plus sign indicates the Galactic Center. In the southern sky, the most significant point is $301.15^{\circ}$ r.a. and $-34.15^{\circ}$ dec, indicated with a box. The map in the northern sky is identical to ~\cite{4YrPSPaper}.}
\label{fig:MESESkyMap}
\end{figure}

The results from both tests are consistent with the background-only hypothesis. The skymap of pre-trial $p$-values is shown in Figure~\ref{fig:MESESkyMap}. In the southern sky, the skymap appears sparser because we observe only 549 starting track events. In the joint likelihood fit, the much lower background of the starting track sample suppresses fluctuations seen in the through-going sample, which has ~100 times greater background.

The location of the most significant pre-trial $p$-value is $301.15^{\circ}$ r.a. and $-34.15^{\circ}$ dec., where one starting track event is coincident with a small excess of events from the through-going muon sample. The likelihood fits 6.97 signal events with an $E^{-2.15}$ energy spectrum, resulting in a pre-trial $p$-value of $9.3 \times 10^{-5}$. Accounting for the trial factor associated with searching every location in the southern sky, the post-trial $p$-value is 0.97.

\begin{figure}
\includegraphics[width=\linewidth]{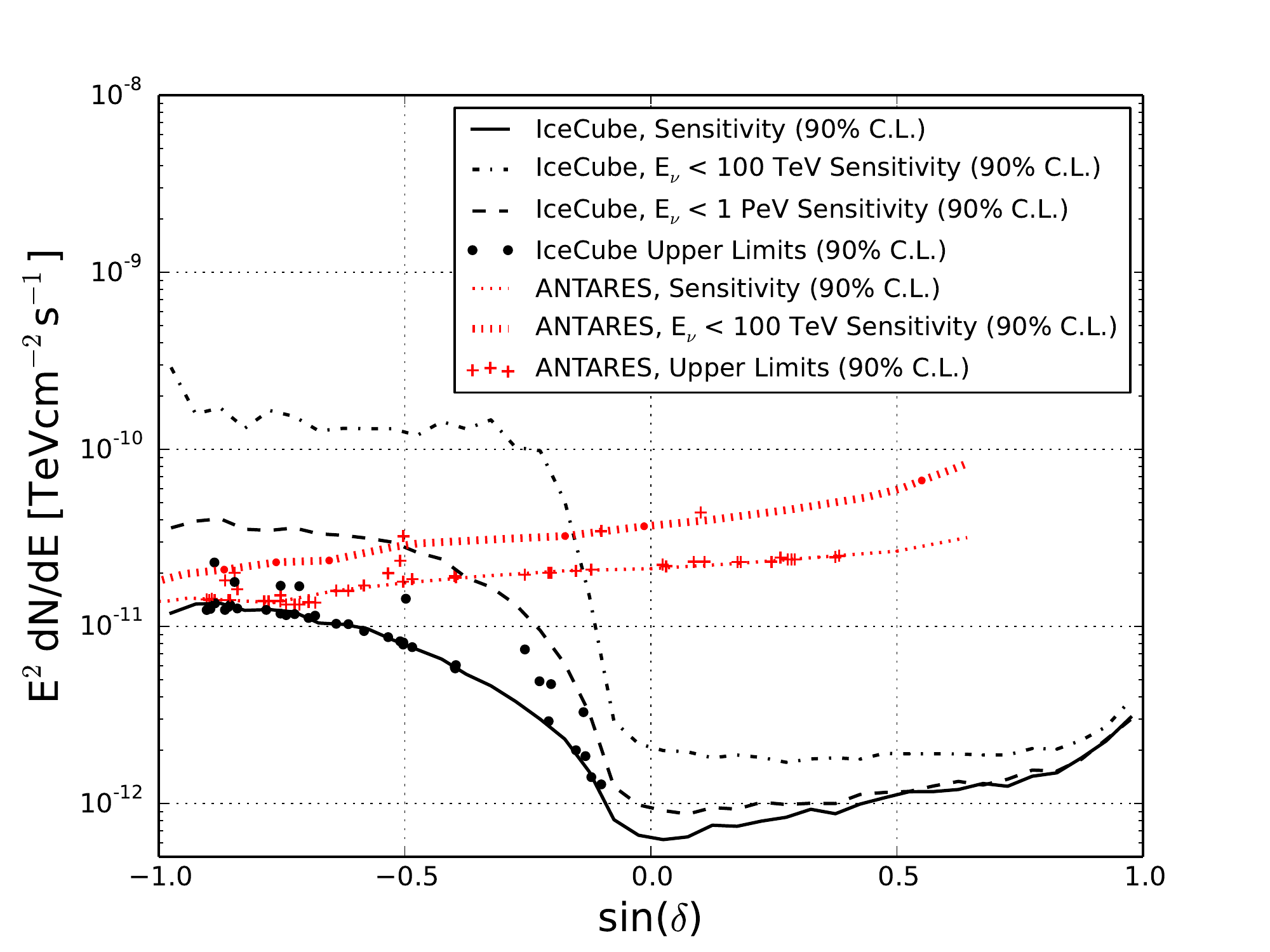}
\caption[Muon neutrino upper limits and sensitivities as a function of declination for the starting track analysis]{Muon neutrino upper limits (90\% C.L.) evaluated for 38 sources (dots), for the three-year starting track sample combined with the four-year throughgoing muon sample. The solid black line is the median 90\% C.L. upper limit or sensitivity for a point source with an unbroken $E^{-2}$ spectrum. The sensitivities to $E^{-2}$ spectra ending (with a sharp cutoff) at \unit[1]{PeV} and \unit[100]{TeV} are shown in the black dashed and dash-dotted lines, respectively. The ANTARES (1338 days livetime) upper limits and sensitivities for two spectral hypotheses are shown in red~\citep{ANTARESPS}.}
  \label{fig:MESEPSSens}
\end{figure}

Results for the {\it a priori} source list search are shown in Table~\ref{tab:MESESourceList}. The source with the strongest excess is the Galactic x-ray binary LS 5039, with a pre-trial $p$-value of 0.10. Accounting for the trial factor associated with the 38 sources on the list, the post-trial $p$-value is 0.76. Figure~\ref{fig:MESEPSSens} shows the 90\% confidence level upper limits and sensitivity on the $E^{-2}$ neutrino flux from each source at its corresponding declination. Also shown is the analysis sensitivity, along with results from ANTARES~\citep{ANTARESPS}.


Since the background is estimated by scrambling the detector data in right ascension, the $p$-values are independent of theoretical uncertainties in the fluxes of atmospheric backgrounds as well as systematic uncertainties in the detector simulation. However, upper limits and analysis sensitivities are calculated by simulating the detector response to astrophysical neutrinos and are subject to these uncertainties. Using detector simulation, the systematic uncertainties in the optical properties of the ice and the efficiency of the optical modules were estimated to have a 16\% and 15\% effect on the analysis sensitivity, respectively.  Summing these uncorrelated errors in quadrature gives a 22\% overall systematic uncertainty on the quoted sensitivities and upper limits, similar to that found in~\cite{4YrPSPaper}.

\begin{deluxetable}{lrrrrrrr} 
\tablecolumns{8} 
\tablewidth{0pc} 
\tablecaption{Results of the {\it a priori} source list} 
\tablehead{ 
\colhead{Category} & \colhead{Source} & \colhead{RA ($^{\circ}$)} & \colhead{Dec ($^{\circ}$)} & \colhead{${\hat n}_s$} & \colhead{${\hat \gamma}$} & \colhead{$p$-value} & \colhead{$\Phi_{\nu_{\mu} + \bar{\nu}_{\mu}}^{90\%}$}
}
\startdata
\cutinhead{\bf Galactic Sources}
SNR    &                W28 & 270.43 & -23.34 &0.0& -- & -- & 6.0\\
&    RX J1713.7-3946 & 258.25 & -39.75 &0.0& -- & -- & 10.4\\
&    RX J0852.0-4622 & 133.0  & -46.37 &0.0& -- & -- & 11.7\\
&             RCW 86 & 220.68 & -62.48 &2.3&2.0&0.28& 23.0\\
\hline
XB/mqso    &            LS 5039 & 276.56 & -14.83 &3.0&2.9&0.10& 7.4\\
&           GX 339-4 & 255.7  & -48.79 &1.7&1.9&0.41& 17.0\\
&            Cir X-1 & 230.17 & -57.17 &0.0& -- & -- & 12.6\\
\hline
Pulsar/PWN   &             Vela X & 128.75 & -45.6  &0.8&2.9&0.44& 16.9\\
&     HESS J1632-478 & 248.04 & -47.82 &0.0& -- & -- & 11.6\\
&     HESS J1616-508 & 243.78 & -51.40 &0.0& -- & -- & 12.4\\
&     HESS J1023-575 & 155.83 & -57.76 &0.5&1.7&0.46& 17.8\\
&          MSH 15-52 & 228.53 & -59.16 &0.0& -- & -- & 12.9\\
&     HESS J1303-631 & 195.74 & -63.52 &0.0& -- & -- & 12.6\\
&     PSR B1259-63 & 195.74 & -63.52 &0.0& -- & -- & 12.6\\
&     HESS J1356-645 & 209.0  & -64.5  &0.0& -- & -- & 12.4\\
   \hline
Galactic  &             Sgr A* & 266.42 & -29.01 &0.0& -- & -- & 7.6\\
Center  &    & &  &  & & & \\
\hline
Not  &     HESS J1834-087 & 278.69 &  -8.76 &0.0& -- & --& 2.0\\
Identified &     HESS J1741-302 & 265.25 & -30.2  &0.0& -- & -- & 8.1\\
&     HESS J1503-582 & 226.46 & -58.74 &0.0& -- & -- & 13.2\\
&     HESS J1507-622 & 226.72 & -62.34 &0.0& -- & -- & 13.5\\
\cutinhead{\bf Extragalactic Sources}
BL Lac &       1ES 0347-121 &  57.35 & -11.99 &0.0& -- & -- & 2.9\\
&       1ES 1101-232 & 165.91 & -23.49 &0.0&--&--& 5.9\\
&       PKS 2155-304 & 329.72 & -30.22 &0.0& -- & -- & 7.9\\
&         H 2356-309 & 359.78 & -30.63 &0.0& -- & -- & 8.2\\
&       PKS 0548-322 &  87.67 & -32.27 &0.0& -- & -- & 8.7\\
&       PKS 0426-380 &  67.17 & -37.93 &0.0& -- & -- & 10.3\\
&       PKS 0537-441 &  84.71 & -44.08 &0.0& -- & -- & 11.2\\
&       PKS 2005-489 & 302.37 & -48.82 &0.0& -- & -- & 11.8\\
\hline
FSRQ   &              3C279 & 194.05 &  -5.79 &0.0& -- & -- & 1.3\\
&     HESS J1837-069 & 279.41 &  -6.95 &0.0& -- & -- & 1.4\\
&       QSO 2022-077 & 306.42 &  -7.64 &0.9&1.9&0.46& 1.9\\
&       PKS 1406-076 & 212.24 &  -7.87 &6.3&2.6&0.10& 3.3\\
&       PKS 0727-11 & 112.58 & -11.7  &4.7&3.4&0.18& 4.7\\
&       QSO 1730-130 & 263.26 & -13.08 &2.4&3.9&0.28& 4.9\\
&       PKS 0454-234 &  74.27 & -23.43 &0.0& -- & -- & 5.8\\
&       PKS 1622-297 & 246.53 & -29.86 &4.1&2.5&0.19& 14.3\\
&       PKS 1454-354 & 224.36 & -35.65 &0.0& -- & -- & 9.4\\
\hline
Radio &  Cen A & 201.37 & -43.02 &0.0& -- & -- & 11.5\\
Galaxies & & & & & & &\\
\hline
Seyfert & ESO 139-G12 & 264.41 & -59.94 &0.0& -- & -- & 12.4\\
\hline

\enddata
\tablecomments{Galactic sources are grouped according to their classification as High-Mass X-ray binaries or micro-quasars (HMXB/mqso), SNRs, Pulsar Wind Nebulas (PWNs), star formation regions and unidentified sources. Extragalactic sources are grouped according to their classification as BL Lac objects, Radio Galaxies, Flat-Spectrum Radio Quasars (FSRQ) and Starburst galaxies. The $p$-value is the (pre-trial) probability for each source direction to have a more significant excess due to background fluctuations. The $\hat{n}_{S}$ and $\hat{\gamma}$ columns give the best-fit number of signal events and spectral index of a power-law spectrum. When $\hat{n}_{S} = 0$, no $p$-value or $\hat{\gamma}$ are reported. The last column shows the $\nu_\mu + \bar{\nu}_{\mu}$ flux normalization classical upper limits~\citep{Neyman} for an unbroken $E^{-2}$ flux in units of $10^{-12}$\,TeV$^{-1}$ cm$^{-2}$s$^{-1}$.}
\label{tab:MESESourceList}
\end{deluxetable} 

\clearpage

\section{Discussion}

\subsection{Constraints on Neutrino Emission Models}

While Figure~\ref{fig:MESEPSSens} shows upper limits for the $E^{-2}$ neutrino flux from a variety of sources, many models predict fluxes with complex spectra. For example, the Galactic Center has been the subject of recent neutrino emission models and discussion~\citep{HalzenGC,RazzaqueGC,AnchoReview,SupanitskyGC}. This is spurred not only by electromagnetic observations~\citep{MorrisSerabyn,CrockerBubbles1,TovaGC}, but also by the observation of high-energy astrophysical neutrinos~\citep{HESEPRL}. The analysis presented here places constraints on the most optimistic possibilities. \citet{HalzenGC} calculated that a point source with an unbroken $E^{-2}$ power-law energy spectrum and flux normalization of $6 \times 10^{-11}$ $\unit{TeV}^{-1}\unit{cm}^{-2}\unit{s}^{-1}$ at \unit[1]{TeV} could be responsible for the high-energy neutrino events near the Galactic Center. This flux is eight times greater than the Galactic Center upper limit presented here. Even if the $E^{-2}$ neutrino spectrum cuts off at \unit[1]{PeV}, it is still excluded at 90\% by this analysis, although an extended source could evade detection.  On the other hand, the majority of models based on gamma-ray observations for this and other Galactic sources predict lower fluxes, many with spectral cutoffs below \unit[100]{TeV}~\citep{KistlerBeacom,KappesEtAl,FoxTeVUnID}. Model tests of these fluxes lead to upper limits that are generally a factor of 10 or more above predictions. 

\subsection{\textbf{\textit{A posteriori}} investigation of the hottest spot in the skymap}

While the hottest spot in the skymap was consistent with a background fluctuation, a single starting track event was the main contributor to the significance at this location. The event appears to be a high-energy $\nu_{\mu}$ interacting inside the detector volume. The reconstructed vertex of this event is \unit[286]{m} inside the detector, and it passes through three layers of PMTs that observe no photons. This event deposited \unit[84]{TeV} (electromagnetic equivalent) inside the detector, which sets a strict lower bound on the neutrino eneryg \citep{EnergyRecoPaper}. When reconstructed as a through-going track, the reconstructed muon energy proxy for this event is \unit[124]{TeV}, the highest in the starting event sample. Its reconstructed direction is $301.7^{\circ}$ r.a. and $-34.3^{\circ}$ dec. with an estimated angular uncertainty of $0.6^{\circ}$, and its arrival time in MJD is 56093.1796492. This event was not found in \citet{HESEPRL} because its deposited charge is below the threshold for that analysis.

At this energy and zenith angle, the expected atmospheric $\nu_{\mu}$ background is greatly suppressed by the atmospheric neutrino self-veto effect~\citep{JakobVetoPaper,SchonertVetoPaper}. In the {\it a posteriori} analysis described here, we characterize the distribution of expected background events in zenith and energy to estimate the significance of having at least one event similar to the one observed in the dataset.

\begin{figure}[t!]
\includegraphics[width=0.495\linewidth]{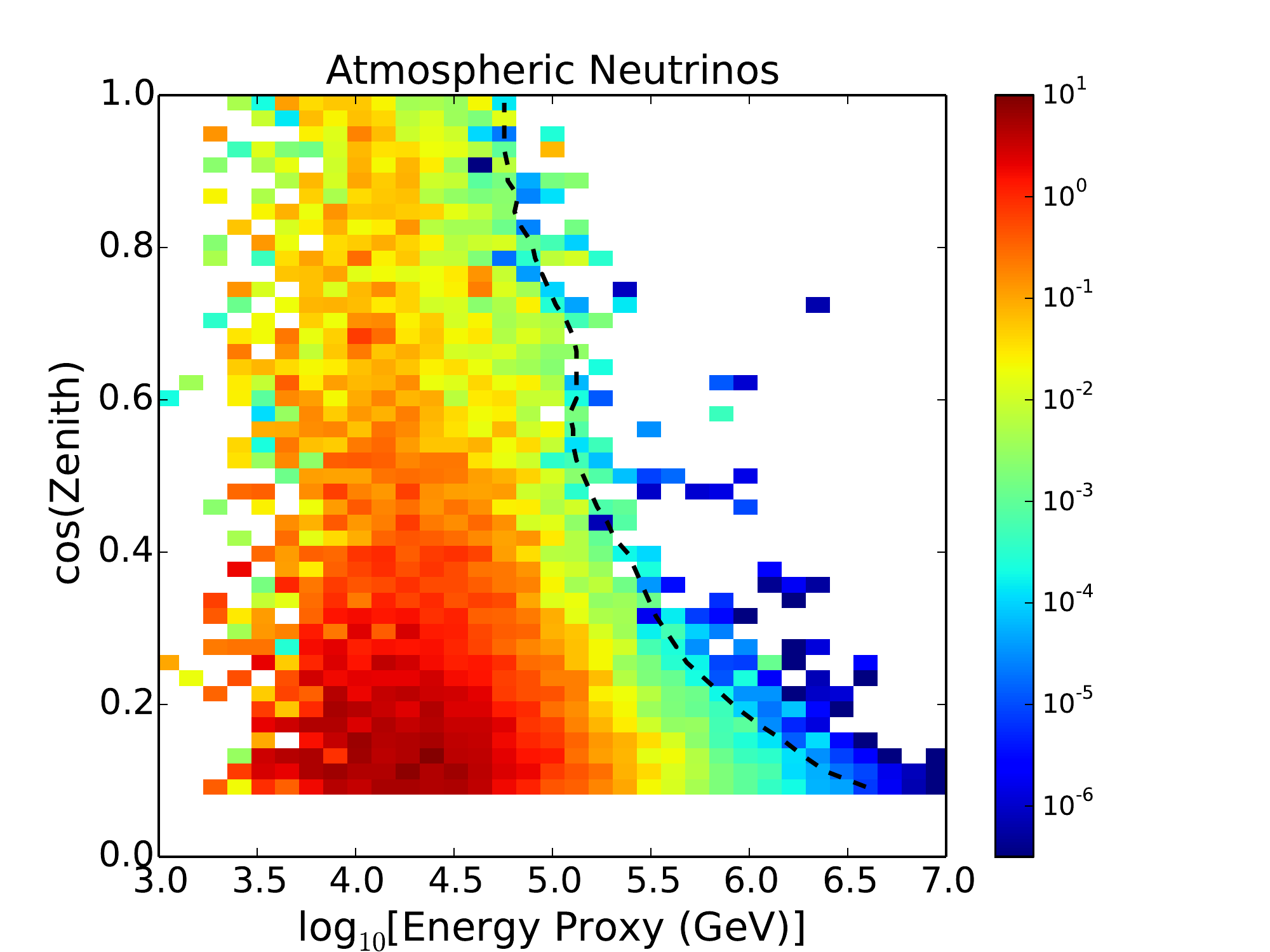}
\includegraphics[width=0.495\linewidth]{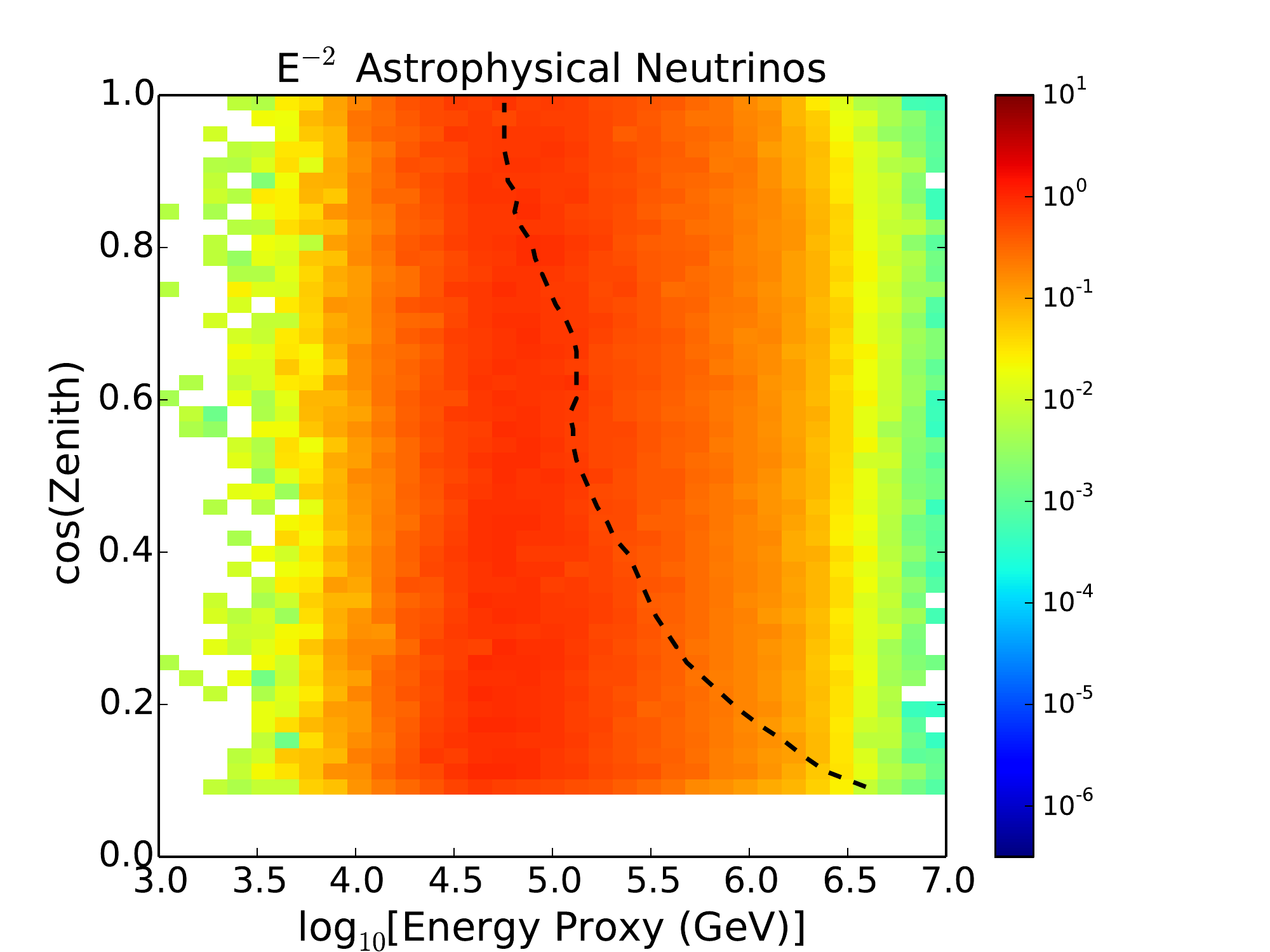}
\caption[Atmospheric $\nu_{\mu}$ background and astrophysical signal PDFs]{Atmospheric $\nu_{\mu}$ background and astrophysical signal PDFs. The left plot shows the PDF as a function of reconstructed zenith angle and reconstructed energy for atmospheric neutrinos following models from \cite{Honda2006,EnbergPrompt,GaisserH3a,AnneDiffuse}. This includes the self-veto probability from \cite{JakobVetoPaper} for each event. The right plot shows the PDF for $E^{-2}$ astrophysical neutrinos. The likelihood ratio between the signal and background hypothesis is calculated for the observed event's energy and zenith, and the dashed line denotes the zenith-energy contour with this likelihood ratio. Events to the right of this contour are more ``signal-like" than the observed event. }
  \label{fig:SigBkgPDF}
\end{figure}

To calculate the total expected background rate, we generate a PDF for astrophysical $\nu_{\mu}$ signal and atmospheric $\nu_{\mu}$ background as a function of reconstructed zenith and energy.  We then take the ratio of these two PDFs to determine the signal to background likelihood ratio for this event's zenith and energy (Fig.~\ref{fig:SigBkgPDF}). The total atmospheric $\nu_{\mu}$ background rate is then the rate of simulated events with a higher likelihood ratio. The total rate of such atmospheric events is 0.0022 in three years of livetime. This event therefore represents a $2.8\sigma$ deviation from the hypothesis that it is an atmospheric background.

Atmospheric muons are a less likely candidate. Muons were simulated with directions, energies, and positions similar to the observed event. The expected rate of atmospheric muons that appear similar to the observed event is estimated to be less than 0.0001 events in three years, an order of magnitude lower than the atmospheric neutrino background.

While this event cannot be easily explained with the background hypothesis, it is consistent with an astrophysical signal hypothesis. The best-fit astrophysical neutrino flux measured in~\citet{HESEPRL} would produce 4.1 starting track events in this dataset, including 1.0 events with a muon energy proxy $\geq \unit[124]{TeV}$. Simulated signal events with similar reconstructed energies had primary neutrino energies of a few hundred \unit{TeV}, most of which is carried out of the detector by the muon.

It is important to note the $2.8\sigma$ significance represents the chance probability that this event originates from an atmospheric flux. It does not include information about its spatial clustering with other events or its correlation with known astrophysical sources. As there is no significant evidence for clustering nor correlation, this event does not represent the identification of a neutrino point source.

\section{Conclusion}

A new event selection technique presented here extends IceCube's sensitivity to southern sky point sources in the energy region below 100 \unit{TeV}. No indication of a statistically significant source was found. Future analyses will probe fainter and lower energy sources in the southern sky by extending this technique further. Joint analyses with ANTARES~\citep{ICANTARESJoint} and future kilometer- and multi-kilometer-scale telescopes~\citep{KM3NeTTDR, IceCubeGen2} will also push sensitivities in this region of the sky. 

\acknowledgments

We acknowledge the support from the following agencies:
U.S. National Science Foundation-Office of Polar Programs,
U.S. National Science Foundation-Physics Division,
University of Wisconsin Alumni Research Foundation,
the Grid Laboratory Of Wisconsin (GLOW) grid infrastructure at the University of Wisconsin - Madison, the Open Science Grid (OSG) grid infrastructure;
U.S. Department of Energy, and National Energy Research Scientific Computing Center,
the Louisiana Optical Network Initiative (LONI) grid computing resources;
Natural Sciences and Engineering Research Council of Canada,
WestGrid and Compute/Calcul Canada;
Swedish Research Council,
Swedish Polar Research Secretariat,
Swedish National Infrastructure for Computing (SNIC),
and Knut and Alice Wallenberg Foundation, Sweden;
German Ministry for Education and Research (BMBF),
Deutsche Forschungsgemeinschaft (DFG),
Helmholtz Alliance for Astroparticle Physics (HAP),
Research Department of Plasmas with Complex Interactions (Bochum), Germany;
Fund for Scientific Research (FNRS-FWO),
FWO Odysseus programme,
Flanders Institute to encourage scientific and technological research in industry (IWT),
Belgian Federal Science Policy Office (Belspo);
University of Oxford, United Kingdom;
Marsden Fund, New Zealand;
Australian Research Council;
Japan Society for Promotion of Science (JSPS);
the Swiss National Science Foundation (SNSF), Switzerland;
National Research Foundation of Korea (NRF);
Villum Fonden, Danish National Research Foundation (DNRF), Denmark

\end{document}